%%%%%%%%%%%%%%%%%%%%%%%%%%%%%%%%%%%%%%%%%%%%%%%%%%%%%%%%%%%%%%%%
%%%%%%%%  Graded Parafermions: spectrum and bases 
%%%%%%%%  P. Jacob and P. Mathieu
%%%%%%%%%%%%%%%%%%%%%%%%%%%%%%%%%%%%%%%%%%%%%%%%%%%%%%%%%%%%%%%%

%\magnification1100

\input harvmac.tex

%\draft

%-------------------------------------------------------------------------------
% contractions de Wick
%
\def\ubrackfill#1{$\mathsurround=0pt
	\kern2.5pt\vrule depth#1\leaders\hrule\hfill\vrule depth#1\kern2.5pt$}
\def\contract#1{\mathop{\vbox{\ialign{##\crcr\noalign{\kern3pt}
	\ubrackfill{3pt}\crcr\noalign{\kern3pt\nointerlineskip}
	$\hfil\displaystyle{#1}\hfil$\crcr}}}\limits
}

\def\ubrack#1{$\mathsurround=0pt
	\vrule depth#1\leaders\hrule\hfill\vrule depth#1$}
\def\dbrack#1{$\mathsurround=0pt
	\vrule height#1\leaders\hrule\hfill\vrule height#1$}
\def\ucontract#1#2{\mathop{\vbox{\ialign{##\crcr\noalign{\kern 4pt}
	\ubrack{#2}\crcr\noalign{\kern 4pt\nointerlineskip}
	$\hskip #1\relax$\crcr}}}\limits
}
\def\dcontract#1#2{\mathop{\vbox{\ialign{##\crcr
	$\hskip #1\relax$\crcr\noalign{\kern0pt}
	\dbrack{#2}\crcr\noalign{\kern0pt\nointerlineskip}
	}}}\limits
}

\def\ucont#1#2#3{^{\kern-#3\ucontract{#1}{#2}\kern #3\kern-#1}}
\def\dcont#1#2#3{_{\kern-#3\dcontract{#1}{#2}\kern #3\kern-#1}}

%=================================================================
% MACROS

%* on my mac I do not have these: so start by decommenting
%* the following set and comment mine

% TABLEAU NOIR (BLACKBOARD)
\font\tenmsy=msbm10
\font\sevenmsy=msbm10 at 7pt
\font\fivemsy=msbm10 at 5pt
\newfam\msyfam % family 11
\textfont\msyfam=\tenmsy
\scriptfont\msyfam=\sevenmsy
\scriptscriptfont\msyfam=\fivemsy
\def\blackB{\fam\msyfam\tenmsy}
\def\Z{{\blackB Z}}

\def\II{{\blackB I}}

\let\R\rangle

\def\vp{\tilde{\varphi}}
\let\da\dagger

\def\frac#1#2{{\textstyle{#1\over #2}}}

% alignements multiples
\def\eqalignD#1{
\vcenter{\openup1\jot\halign{
\hfil$\displaystyle{##}$~&
$\displaystyle{##}$\hfil~&
$\displaystyle{##}$\hfil\cr
#1}}
}
\def\eqalignT#1{
\vcenter{\openup1\jot\halign{
\hfil$\displaystyle{##}$~&
$\displaystyle{##}$\hfil~&
$\displaystyle{##}$\hfil~&
$\displaystyle{##}$\hfil\cr
#1}}
}

\def\text#1{\quad\hbox{#1}\quad}
\def\gh{\hat{g}}

\def\e{\epsilon}

\def\A{{\cal{A}}}
\def\B{{\cal{B }}}
\def\J{\widehat {J}}

\def\y{{\infty}}

\def\gh{{\widehat g}}
\def\rw{\rightarrow}

\def\O{{\cal O}}

\def\R{\rangle}

\def\su{\widehat{su}}
\def\osp{\widehat{osp}}

%\def\E1{E_{1}}

% Equations (overrides harvmac's equation macros)
\newcount\eqnum
\eqnum=0
\def\eq{\eqno(\secsym\the\meqno)\global\advance\meqno by1}
\def\eqlabel#1{{\xdef#1{\secsym\the\meqno}}\eq }

% References (overrides harvmac's reference macros)
\newwrite\refs
\def\startreferences{
 \immediate\openout\refs=references
 \immediate\write\refs{\baselineskip=14pt \parindent=16pt \parskip=2pt}
}
\startreferences

\refno=0
\def\aref#1{\global\advance\refno by1
 \immediate\write\refs{\noexpand\item{\the\refno.}#1\hfil\par}}
\def\ref#1{\aref{#1}\the\refno}
\def\refname#1{\xdef#1{\the\refno}}

\newcount\exno
\exno=0
\def\Ex{\global\advance\exno by1{\noindent\sl Example \the\exno:

\nobreak\par\nobreak}}

\parskip=6pt

\overfullrule=0mm

%%%%%%%%%%%%%%%%%%%%%%%%%%%%
\def\frac#1#2{{#1 \over #2}}

\def\uh{{\widehat u}}
\def\rw{{\rightarrow}}

\def\Zt{{\widetilde\Z}}

% References (overrides harvmac's reference macros)
\newwrite\refs
\def\startreferences{
 \immediate\openout\refs=references
 \immediate\write\refs{\baselineskip=14pt \parindent=16pt \parskip=2pt}
}
\startreferences

\refno=0
\def\aref#1{\global\advance\refno by1
 \immediate\write\refs{\noexpand\item{\the\refno.}#1\hfil\par}}
\def\ref#1{\aref{#1}\the\refno}
\def\refname#1{\xdef#1{\the\refno}}

%===============================================================================

% PAGE TITRE
\Title{\vbox{\baselineskip12pt
\hbox{ }}}
{\vbox {\centerline{ Graded parafermions: standard and quasi-particle bases}
\bigskip
%\centerline{ and generalized
%  commutation relations}
}}

\smallskip
\centerline{ P. Jacob and P. Mathieu
% \foot{Work supported by NSERC (Canada) and FCAR (Qu\'ebec) }
% \foot{Corresponding author (PM): tel:
% 418-656-3416; fax: 418-656-2040 }
}

\smallskip\centerline{ \it D\'epartement de
physique,} \smallskip\centerline{\it Universit\'e Laval,}
\smallskip\centerline{ \it Qu\'ebec, Canada G1K 7P4}
\smallskip\centerline{(pjacob@phy.ulaval.ca, pmathieu@phy.ulaval.ca)}
\vskip .2in
\bigskip
%\bigskip
%\bigskip
%\bigskip
\bigskip
\centerline{\bf Abstract}
\bigskip
\noindent

Two bases of states  are presented for modules of the
graded parafermionic conformal field
theory associated to the coset
$\osp(1,2)_k/\uh(1)$.
The first one is  formulated in terms of the two fundamental
(i.e., lowest dimensional) parafermionic modes. In that basis, one can identify the completely
reducible representations, i.e., those  whose modules contain an infinite number of singular
vectors; the explicit form of these vectors is also given. 
 The second basis is a
quasi-particle basis, determined in terms of a modified version of the $\Z_{2k}$  exclusion principle.
A novel feature of this model is that none of its bases are fully ordered and this reflects a hidden
structural $\Z_3$ exclusion principle.

%Classification numbers: 11.10.-z, 02.20+bm

%Keywords: conformal field theory, parafermions, 
%$Z_N$ models; character; singular vectors.
\Date{09/01\ }
%\ (hepth@xxx/0201156)}

%==============================================================================

%%%%% ref a Fring - voir Melzer susy case...
\newsec{Introduction}

Conformal fields theories are normally described in terms of the representation theory of the extended
conformal algebra.  However, they can also be described in terms of some quasi-particle excitations.
This  provides a quite interesting complementary formulation, interesting both from the physical and the
mathematical points of view. 

On the mathematical side, a quasi-particle basis leads to fermionic-sum expressions
for the characters, with an underlying Andrews-Gordon-type combinatorics. Such characters are quite distinct
in their structure from the standard bosonic-type characters derived from a Verma module description  with
singular-vector subtractions.\foot{The qualitative `bosonic' and `fermionic' for character-sum
representations are explained in more details in [\ref{R. Kedem, T.R.
Klassen, B. M. McCoy and E. Melzer, Phys. Lett. {\bf B304} (1993) 263.}\refname\KKMM, \ref{E.
Melzer, Int. J. Mod. Phys. {\bf A9} (1994) 133.}\refname\Mel] and in the introduction of [\ref{P. Jacob and P.
Mathieu, {\it Parafermionic quasi-particle basis and fermionic-type characters }, Nucl. Phys. B - to
appear, hep-th/0108063}\refname\JMb], where
the origin of the fermionic sums in CFT is also traced back.}
Constructing both the bosonic and fermionic characters of a given theory provides a field theoretical
derivation of some highly nontrivial mathematical identities.

On the other hand, a description of a CFT in terms of quasi-particles is manifestly closer to the usual
quantum field theoretical framework. It can be viewed as a reformulation of the CFT in
preparation for an integrable perturbation to be described, off-critically, in terms of the
corresponding massive excitations [\KKMM].

The physical interpretation of the  quasi-particle excitations is not always transparent however. This point
can be neatly illustrated with the particular case of the Lepowsky-Primc characters [\ref{J. Lepowsky and M.
Primc,  Contemporary Mathematics {\bf 46} AMS, Providence, 1985.}\refname\LP], the prototype of  fermionic-sum
representations of CFT characters.  These characters count the number of states at each level for a 
generic basis spanned by some creation operators ${\cal C}_n$ -- subject to a  given (generalized) commutation
relation --
% \foot{By generalized commutation lreations, we refer to commuation
% rules that are written in terms of an infinte sum, such as those pertaining to the $\Z$--algebra
% [\ref{J. Lepowsky and R.L. Wilson, Inv. Math. {\bf 77} (1984) 199.}], the spinon bases [\BLS] or the
% parafermionic algebra ..}
acting on a highest-weight state $|\varphi\R$ and  ordered as
$${\cal C}_{-n_1}{\cal C}_{-n_2}....{\cal C}_{-n_p}|\varphi\R\qquad \qquad n_i\geq n_{i+1}\geq
1\eqlabel\genba$$ up to some boundary condition (i.e., a constraint on the maximal number of 1's), with the
further requirement that
$$ n_i\geq n_{i+k-1}+2\eqlabel\restru$$
This is the core restriction; it characterizes the combinatorics of the Andrews-Gordon identities
that generalize those of Rogers-Ramanujan-Schur.\foot{Note that an equivalent way of 
formulating this basis is
by writing the states (\genba) under the form
$${\prod^{\leftarrow}}_{i\geq 1} ({\cal C}_{-i})^{a_i}|\varphi\R \qquad \quad a_i\geq 0$$
where the arrow indicates that the index $i$ increases toward the left; the constraint (\restru) is
equivalent to imposing 
$a_i+a_{i+1}< k$.}   The above restriction is rooted in the existence of
one relation at the level of fields, a  relation that is model dependent.
  The original derivation of
Lepowsky-Primc applies to the coset
$\su(2)_k/\uh(1)$ and the relation that leads to the restriction rule is (cf. [\LP], prop. 5.5)
$$e(z)^{k+1}= f(z)^{k+1}=0\eqlabel\nul$$ 
More precisely, $e(z)^{k+1}$ or $f(z)^{k+1}$,  applied on any
field associated to a state in an integrable
$\su(2)_k$ module, vanishes.  This null-field condition  
leads to a linear relation among the spanning states at a given grade, whose effect is captured by the
restriction rule (\restru).\foot{Here
$e$ and
$f$ are the `raising' and `lowering'
$\su(2)$ currents in the Chevalley basis. This relation is easily shown for
$k=1$ and the result is lifted to general $k$ by tensor product.  The same relation  also appears in
the
$\su(2)_k$ fermionic bases constructed in
[\ref{B.L. Feigin and A.V. Stoyanovski, RIMS 982, hep-th/ Func, Anal. Appl. {\bf 28} (1994) 68; B.L.
Feigin and T. Miwa, in {\it Statistical physics at the eve of the 21st century}, 
Series on Adv. in Stat. Mech., vol 14, ed. by M.T Batchelor and L.T. White, World Scientific, 1999; B.
Feigin, R. Kedem, S. Loktev, T. Miwa, E. Mukhin, Transformation Groups, {\bf 6 }  (2001) 25.}].}

%%% check the modification...
Now these characters, in a slightly modified form, also turn out to provide
a fermionic-sum representation for the Virasoro minimal models
${\cal M}(2,2k+1)$ [\ref{V. Kac, {\it Modular invariance in mathematics and  physics}, address at the
centennial of the AMS, (1988).}\refname\Kac,  \ref{B.L. Feigin, T. Nakanishi  and H. Ooguri, Int. J.
Mod. Phys. {\bf A7} Suppl. {\bf 1A} (1992) 217.}\refname\FNO]. In this context, the restriction rule is
rooted in the null field condition
$T(z)^{k}+\cdots$ (the power of $T$ being suitably normal ordered and the dots
stand for a differential polynomial in $T$ of degree $2k$).  This is the field version
of the non-trivial vacuum singular vector, the one at level
$2k$ [\FNO, \ref{B.L. Feigin and E. Frenkel, Adv. Sov. Math. {\bf 16} (1993) 139.}\refname\FF].
Although the mathematical structure of the irreducible modules of the $\su(2)_k/\uh(1)$ and the
${\cal M}(2,2k+1)$ models are similar -- i.e., there exists a linear relation that depends upon $k$), their physical
excitations are quite different.\foot{Another example exhibiting a similar mathematical structure but with a
quite different source  for the constraining linear relation, is provided by the spinon description of  the
$\su(k)_1$ WZW models [\ref{D. Bernard,  V. Pasquier and
D. Serban, Nucl. Phys. {\bf B 428} (1994) 612.}\refname\serban, \ref{ P. Bouwknegt. A.A. Ludwig and K. 
Schoutens, Phys. Lett. {\bf B 338} (1994) 448; 
{\bf B 359} (1995) 304.}\refname\BLS, \ref{P. Bouwknegt and K. Schoutens, Nucl. Phys. B 482 (1996) 345;
{\it Spinon decomposition and Yangian structure of $(sl_n)$ modules} to appear in  {\it Geometric
Analysis and Lie Theory in Mathematics and Physics}, Lecture Notes Series of the Australian
       Mathematical Society..}\refname\schou]. 
In that case, the  restrictions on 
the ordered spanning states originates from the Yangian symmetry inherited from the finitized version of
the theory,  the Haldane-Shastry spin chain  [\ref{F.D.M. Haldane, Z.N.C. Ha, J.C. Talstra, D. Bernard
and  V. Pasquier, Phys. Rev. Lett. {\bf 69} (1992)  2021.}\refname\pasquier]; it also leads to 
restriction rules incorporating (\restru) - cf. the first reference in [\schou], app. A.} 
% The spinon basis can be obtained either 
% via this symmetry or directy
% from the  generalized commutation 
% relations between the spinon modes (as in the second reference in
% [\schou]). 
In fact, the Virasoro modes (in terms of which the ${\cal M}(2,2k+1)$ space of states are described) can even not
be regarded as physical quasi-particles.  Actually, even in the first case, the
physical meaning of the excitations is not clear at once.  However, their proper interpretation is
unveiled  in the observation that
$\su(2)_k/\uh(1)$ is a coset representation of the $\Z_k$  parafermionic theory [\ref{A.B.
Zamolodchikov and V.A. Fateev, Sov. Phys. JETP {\bf 43} (1985) 215.}\refname\ZFa].

The  parafermionic quasi-particle basis has been derived in [\JMb] using solely
the defining generalized commutation relations. The mere $\Z_k$ invariance  -- that is, $(\psi_1)^k \sim
\II$, where $\psi_1$ is the fundamental parafermion --, has then been identified as the source for the
restriction rule (\restru).  Therefore, the mathematical structure underlying the restrictions
rules is, in some sense, built in the theory, in
the $\Z_k$ invariance of the OPE's.  Moreover, in this context, the quasi-particle excitations have a natural
physical interpretation: these are the modes of the  parafermionic field $\psi_1$, say $\A$, which are subject
to a
$\Z_k$ exclusion principle, which takes the following form:  all ordered sequences
of
$\A$ operators that contain any one of the following $k$-string
$$(\A_{-(n+1)})^{k-i}(\A_{-n})^{i}\qquad (i=0,\cdots , k-1)\eqlabel\parexcl$$
must be forbidden. Excluding the
states (\parexcl) amounts to enforce (\restru).\foot{The interpretation of this condition in terms of
an exclusion principle goes back to [\ref{J. Lepowsky and R.L. Wilson, Proc. Nat. Acad. Sci. USA {\bf
78} (1981) 7254.}\refname\LW].  A more recent discussion can be found in [\ref{C. Dong and J.
Lepowsky {\it Generalized vertex algebras and relative vertex operators}, Birkha\"user, 1993.}\refname\DL]. 
For another formulation of
the generalized  exclusion principle as a constraint on a basis of states, see [\ref{K.  Schoutens, Phys.
 Rev. Lett. {\bf 79} (1997) 2608; P. Bouwknegt. and K. 
 Schoutens, Nucl. Phys. 
 {\bf B 547} (1999) 501.}\refname\bouschou]. Also, the connection between such type of  exclusion
principle and the Haldane's generalization [\ref{F.D.M. Haldane,  Phys. Rev. Lett. {\bf 67} (1991) 
937.}\refname\hal] is clarified in [\ref{A. Berkovich and B.M. McCoy, 
 in {\it Statistical physics at the eve of the 21st century}, 
Series on Adv. in Stat. Mech., vol 14, ed. by M.T Batchelor and L.T. White, World Scientific, 1999.}].}

These observations suggest that the parafermionic  approach followed in [\JMb], in addition to remain close
to the physics, would also be the simplest one for deriving quasi-particle bases for generalized cosets
of the form
${\gh}_k/\uh(1)^r$, $r$ being the rank of $\gh$ (and, of course, this is the only tool at hand for more
general parafermionic models that do not have such a simple coset structure, such as the models described
 in  app. A of [\ZFa]).

The aim of this paper is to supply a partial evidence for this `contention', by presenting the
derivation of a novel quasi-particle basis, the one pertaining to the graded $\Z_k$ parafermions
related to the coset
$\osp(1,2)_k/\uh(1)$. This model has originally been proposed in [\ref{J. M. Camino, A. V. Ramallo and
J. M. Sanchez de Santos, Nucl.Phys. {\bf B530} (1998) 715.}\refname\CRS]. As a side result, we also
display the form of the `standard basis', the structure of the singular vectors and fix the spectrum of
the completely degenerate representations. The characters will be presented in a sequel article.

\newsec{The graded $\Z_k$ model}

\subsec{The OPEs}

The graded version of the $\Z_k$ parafermionic theory, or more precisely, the parafermionic theory
defined by the coset $\osp(1,2)_k/\uh(1)$ where $k$ is a  positive integer, contains the following set
of parafermionic fields:
$\psi_r,\; r=0,\frac12,1,\cdots, k-\frac12$, of dimension
$$h_{\psi_r}= {r(k-r)\over k} +{\e_r\over 2}\eq$$
where $\e_r=0$ if $r$ is integer and 1 otherwise.  Note that $\psi_r=
\psi_{k-r}^\da$ and $\psi_0=\II$. The dimension of the lowest dimensional
parafermion
$\psi_{\frac12}$, is thus $1-1/4k$.  The defining OPEs read
$$\eqalign{ 
\psi_r (z) \,\psi_{s} (w) &\sim {c_{r,s}\over  (z-w)^{2rs/k+\e_r\e_s}}\;
 \psi_{r+s} (w)  \qquad (r+s<k) \cr
 \psi_r (z) \,\psi^\dagger_{s} (w) &\sim { c_{r,k-s}\over 
(z-w)^{2{\rm min}(r,s)-2rs/k +\e_r\e_s} }\;
  \psi_{k+r-s} (w)  \qquad (r+s<k) \cr
\psi_r (z) \,\psi^\dagger_r (w) &\sim {1\over (z-w)^{2r(k-r)/k+\e_r}}
\left[\II+(z-w)^2 \{\e_r \O^{(\frac12)}(w) +(1-\e_r) \O^{(1)}(w) \}+\cdots\right]
\cr} \eqlabel\zkope$$
where $\O^{(\frac12)}$ and $\O^{(1)}$ are two currents of dimension 2, whose difference is proportional to
the energy-momentum tensor [\CRS] 
$$T(z) ={2k\over 2k+3}\left[ \O^{(1)}(z) - \O^{(\frac12)}(z)\right]\eqlabel\tem$$
This expression is forced by associativity and it agrees with the one derived by the coset construction, the
zero-mode of $\O^{(1)} - \O^{(\frac12)}$ being proportional to the $osp(1,2)$ Casimir, up to a $u(1)$
contribution. Associativity also fixes the central charge to be
$$c={-3\over 2k+3}={k\over k+3/2} -1 \eq$$
This is again in agreement with the value calculated from the coset description, which is given by the last
equality: the dimension of $osp(1,2)$ being $1=3-2$ -- three even and two odd generators -- and the dual
Coxeter is
$3/2$. This model will be referred to as  the
$\Zt_k$ parafermionic theory.

The structure constants $c_{r,s}$ are also fixed by associativity. In [\CRS], they have been
obtained mainly by means of a free-field representation. A direct implementation of associativity,
reformulated in a somewhat simplified form, is presented in [\ref{P. Jacob and P. Mathieu, {\it Parafermionic
Jacobi 
 identities,  Sugawara construction and generalized
 commutation relations}, in preparation.}\refname\JMc].

The $\Z_k$ parafermionic algebra being a subalgebra of the $\Zt_k$ model, we can
fix the charge normalization from the former theory:  the charge of $\psi_r$ will then be $2r$ 
(and the charge is defined modulo
$2k$).

\subsec{The generalized commutation relations}

The parafermionic modes are defined as follows:
$$\eqalign{
 \psi_{\frac12}(z)\phi_{q}(0) & = \sum_{m=-\y}^\y
z^{-q/2k-m-1}B_{(1+2q)/4k+m}\, \phi_{q}(0)\cr
\psi_{\frac12}^\dagger (z)\phi_{q}(0)  &= \sum_{m=-\y}^\y
z^{q/2k-m-1} B^\dagger_{(1-2q)/4k+m}\,\phi_{q}(0)\cr
\psi_1(z)\phi_{q}(0) & = \sum_{m=-\y}^\y
z^{-q/k-m-1}A_{(1+q)/k+m}\, \phi_{q}(0)\cr
\psi_1^\dagger (z)\phi_{q}(0)  &= \sum_{m=-\y}^\y
z^{q/k-m-1} A^\dagger_{(1-q)/k+m}\,\phi_{q}(0)\cr}
\eqlabel\modepp$$
We will make use of the convention
$$\eqalignD{
& \B_n |\phi_{q}\R\equiv  B_{n+(1+2q)/4k}|\phi_{q}\R\, ,\qquad
&\B_n^\dagger|\phi_{q}\R\equiv 
B_{n+(1-2q)/4k}^\dagger|\phi_{q}\R\cr
&\A_n |\phi_{q}\R\equiv  A_{n+(1+q)/k}|\phi_{q}\R\, ,\qquad
&\A_n^\dagger|\phi_{q}\R\equiv 
A_{n+(1-q)/k}^\dagger|\phi_{q}\R\cr}\eqlabel\conva$$
that is, the fractional parts will not be written down explicitly.
By evaluating the integral [\ZFa,\CRS] 
% (see also [\JMa, \ref{P.
% Jacob and P. Mathieu, {\it Parafermionic Jacobi 
% identities,  Sugawara construction and generalized
% commutation relations}, in preparation.}\refname\JMc])
$${1\over (2 \pi i)^2}\oint_{C_1} dw\,\oint_{C_2} dz\,
  z^{qr/k+n'}\, w^{qs/k+m'}\,(z-w)^{t}\,
\psi_r (z)\, \psi_s (w)\, \phi_{q} (0)\eqlabel\preint$$with appropriate values of $r, s$ -- recall that
$\psi_p^\da=\psi_{k-p}$ -- and
$t$ (and by redefining $n'$ and $m'$ by suitable shifts), we find the following generalized  commutation
relations [\CRS]:
$$ \eqalign{
& 
 \sum_{l=0}^{\infty} C^{(l)}_{-1-1/2k} \left[ \B_{n-l-1}
\B^\da_{m+l+1} - \B^\da_{m-l} \B_{n+l}
 \right] \phi_{q}(0)  \cr
&\qquad \qquad \qquad =  \left[ {1 \over 2}
({q\over 2k}+n)({q\over 2k}+n-1)\delta_{n+m,0}+ \O^{(\frac12)}_{n+m} \right] \phi_{q}(0)\cr
& \sum_{l=0}^{\infty} C^{(l)}_{-1-2/k} \left[ \A_{n-l-1}
\A^\da_{m+l+1} + \A^\da_{m-l} \A_{n+l}
 \right] \phi_{q}(0)  \cr
& \qquad \qquad \qquad =  \left[ {1 \over 2}
({q\over k}+n)({q\over k}+n-1)\delta_{n+m,0}+ \O^{(1)}_{n+m}\right] \phi_{q}(0) \cr
& 
 \sum_{l=0}^{\infty} C^{(l)}_{-1/k} \left[ \A_{n-l}
\B^\da_{m+l} - \B^\da_{m-l} \A_{n+l}
 \right] \phi_{q}(0)  =   c_{1,k-\frac12}\, \B_{n+m} \phi_{q}(0) \cr
& 
 \sum_{l=0}^{\infty} C^{(l)}_{1/2k} \left[ \B_{n-l}
\B_{m+l} + \B_{m-l} \B_{n+l}
 \right] \phi_{q}(0)  =   c_{\frac12,\frac12}\, \A_{n+m} \phi_{q}(0) \cr
 & 
 \sum_{l=0}^{\infty} C^{(l)}_{1+1/2k} \left[ \B_{n-l}
\B_{m+l} - \B_{m+1-l} \B_{n-1+l}
 \right] \phi_{q}(0)  =   0 \cr
%   & \sum_{l=0}^{\infty} C_{2/k}^{(l)} 
%  \left[ \A_{n-l}\A_{m+l}-\A_{m-l}\A_{n+l} \right]
% \phi_{q}(0) =0 \cr
}\eqlabel\graco$$
together with the dagger version of the last three relations.
$C_t^{(l)}$ stands for the binomial coefficient:
 $$ C_t^{(l)}=  {\Gamma(l-t)\over l!\,
\Gamma(-t)}\eq$$
The fractional part of $t$ is fixed by
the field content of the integrand (i.e., the values of $r,s,q$). By varying its integer part, we
can pick up different contributions in the OPE of $\psi_r$ and $\psi_s$ - cf. the fourth vs the fifth
relations (and other variants will be introduced when required). The relative signs on the left hand side have
various contributions: the parity of the integer power of
$t$, the fermionic/bosonic nature of the two fields $\psi_r,\, \psi_s$ and, of course, the minus sign that
comes from taking the difference between the $|z|>|w|$ and
$|z|<|w|$ contribution.

\subsec{Highest-weight conditions} 

The highest-weight conditions ($q$ being the charge of $\vp_q$, always chosen to be positive) are
$$\eqalign{
& \B_n |\vp_q\R=  \A_n |\vp_q\R= 0\qquad n\geq 0\cr
& \B^\da_{n+1} |\vp_q\R= \A^\da_{n+1} |\vp_q\R =0\qquad n\geq 0\cr}\eq$$ 
Note that the condition on $\B$ implies that on $\A$ (using the fourth relation in (\graco)) but 
the constraint $\A^\da_1|\vp_q\R = 0 $ cannot be deduced from the $\B^\da$ condition.  These
highest-weight conditions imply 
$$ \O^{(\frac12)}_{n} |\vp_q\R = \O^{(1)}_{n}|\vp_q\R  =  L_{n}|\vp_q\R= 0\qquad n\geq
1\eq$$

\newsec{The $\Zt_k$ standard basis}  

 \subsec{Reviewing the $\Z_k$ standard basis}

The aim of this section is to obtain the space of states in highest-weight modules in the
`standard basis', the one that generates the Verma modules.  It is constructed from strings of $\B$ and
$\B^\da$ operators.  

This derivation will rely heavily on results of the $\Z_k$ case. So in this subsection we 
first recall the description of the $\Z_k$ standard basis obtained in [\ref{P. Jacob and
P. Mathieu, Nucl. Phys. {\bf B 587} (2000) 514.}\refname\JMa]. The set of independent states at level
(grade) $s$ in the highest-weight Verma module of
$|\varphi_\ell\R$\foot{The absence of a tilde  distinguishes the usual $\Z_k$ primary fields from the $\Zt_k$
ones.}, the highest-weight conditions being
$$\A_{n} | \varphi_\ell 
\rangle = \A^\dagger_{n+1} | \varphi_\ell \rangle 
=0  \qquad {\rm for}\quad n\geq 0\eqlabel\hiwe$$ 
and with relative  charge
$2r$, is found to be generated by the following states [\JMa]:
$$\A_{-n_1}\A_{-n_2} .. .  \A_{-n_{p}}\A^\da_{-m_1}\A^\da_{-m_2} .. .  \A^\da_{-m_{p'}}
|\varphi_\ell\R\eqlabel\mixs$$
with
$$ p-p'=r\, , \quad n_{i} \geq n_{i+1}\geq 1\,, \qquad m_{i} \geq m_{i+1}\geq 0\,,
\qquad
\sum_{i=1}^{p} n_i+
\sum_{i=1}^{p'} m_i=s\eqlabel\mixst$$
% In that case, the   highest-weight state conditions read
% $$\A_{n} | \varphi_\ell 
% \rangle = \A^\dagger_{n+1} | \varphi_\ell \rangle 
% =0  \qquad {\rm for}\quad n\geq 0\eqlabel\hiwe$$
% In this basis, there is an infinite number of singular 
% vectors and that leads naturally to 
% bosonic-type character formulae, displayed in [\JMa].

\subsec{A bijection between $\Z_k$ $\A$-strings and $\Zt_k$ $\B$-strings}

The argument presented in [\JMa] extends directly to the graded model, with the
$\A,~\A^\da$ modes  replaced by the
$\B,~\B^\da$ ones.\foot{It should be stressed that $\B$ in [\JMa]
refers to $\A^\da$ and not the present $\B$.} 
In particular, the $\B$ and $\B^\da$ strings can be ordered separately
but there is no mixed ordering.

 There is however an important difference in the $\Z_k$ and the $\Zt_k$ ordering 
conditions that hold for
each separate strings of a given type of modes. To study this question, we will establish 
a correspondence
between a $p$-string of $\B$ operators in
the
$\Zt_k$ model and a $p$-string of $\A$ operators in the $\Z_k$ and use this to derive the $\B$-ordering
constraint for the $\A$ one.

This correspondence is suggested by the comparison of the following  two sets of commutation
relations (where we omit the $\phi_q(0)$ term on each side), pertaining respectively to the $\Zt_k$ and the
$\Z_k$ models:
$$\eqalignD{
 &(i)\quad &
 \sum_{l=0}^{\infty} C^{(l)}_{1+1/2k} \left[ \B_{n-l}
\B_{m+l} - \B_{m+1-l} \B_{n-1+l}
 \right]   =   0 \cr
  &(ii)\quad&  
 \sum_{l=0}^{\infty} C^{(l)}_{1/2k} \left[ \B_{n-l}
\B_{m+l} + \B_{m+l} \B_{n+l}
 \right]   =   c_{\frac12,\frac12} \B^{(2)}_{m+n}\cr
  &(iii)\quad & 
 \sum_{l=0}^{\infty} C^{(l)}_{1/k} \left[ \B_{n-l}
\B^{(2)}_{m+l} + \B^{(2)}_{m+l} \B_{n+l}
 \right]   =   0 \cr}\eqlabel\compb$$
and
$$\eqalignD{   
 &(i)\quad &
 \sum_{l=0}^{\infty} C_{2/k}^{(l)} 
  \left[ \A_{n-l}\A_{m+l}-\A_{m-l}\A_{n+l} \right]
 =0 \cr
  &(ii)\quad&
\sum_{l=0}^{\infty} C^{(l)}_{2/k-1} \left[ \A_{n-l}
\A_{m+l} + \A_{m+l-1} \A_{n+l+1}
 \right]   =   c_{1,1} \A^{(2)}_{m+n}\cr
 &(iii)\quad &  
 \sum_{l=0}^{\infty} C^{(l)}_{4/k-1} \left[ \A_{n-l}
\A^{(2)}_{m+l} + \A^{(2)}_{m+l-1} \A_{n+l+1}
 \right]   =   c_{1,2}\A^{(3)}_{n+m} \cr}
\eqlabel\compa$$ 
Here $\B^{(n)}$ stands for the modes of the parafermion
$\psi_{n/2}$, so that $\B^{(2)}\sim\A$, and similarly $\A^{(n)}$ stands for the modes of the parafermion
$\psi_{n}$.
In each case, the second relation is a simple variant of the first one but  taking an extra pole.

Disregarding for the moment the obvious difference  between the right hand side of both versions of 
$(iii)$,   one observes  a manifest structural analogy between both sets of relations. 
However, a closer look reveals  a  dissimilarity  rooted  in
the relative shifts of the indices of the second factor of the left hand side of each relation: it is readily
observed that the mode indices on the second term of each
$\B$ commutator are shifted by
$+1$ and
$-1$ when compared to the
$\A$ terms of the  corresponding commutator.  This means that the $\B$ basis is a priori less constrained.

The ordering constraints are encoded in the relations $(i)$. The relations $(ii)$ is displayed in order to
show that no further information can be obtained by picking out an extra pole in the $\B\B$ commutator: as for
the
$\A\A$ commutator, taking an extra pole brings down another field. This ensures that given the map from $\A$
to $\B$ strings, we will be allowed to extract the $\B$-ordering condition out of the $\A$ one.

Given the above observation concerning the difference in mode indices and given that the $\A$ and $\B$
highest-weight conditions are similar, we define the following map:
$$(n_1',\cdots ,n_p') = (n_1,\cdots ,n_p) + (p-1,p-2,\cdots,1,0)\eqlabel\stair$$  
where we used  the notation 
$$\A_{-n'_1}\cdots \A_{-n'_p} \sim (n_1',\cdots ,n_p')\qquad\qquad \B_{-n_1}\cdots \B_{-n_p} \sim
(n_1,\cdots ,n_p)\eqlabel\appli$$
Therefore, upon subtracting the staircase partition from a general partition describing an ordered
$\A$-string, we thus map the $\A$-ordering condition $n_i'\geq n_{i+1}'$ into a weaker one$$n_i'\geq
n_{i+1}' \qquad \Longrightarrow \qquad n_i\geq n_{i+1}-1\eq$$
holding to $\B$ strings.  The application (\appli) is manifestly a bijection. Note that $(n_1,\cdots,n_p)$
is not a partition.

We thus sees that the minor-looking difference between the type-$(i)$ commutation relations has in fact a
rather dramatic effect on the ordered basis: while the $\A$ index cannot increase from right to left,  in the
$\B$ case, it can increase by one at each single step.

Let us illustrate the effect of this difference for the ordering of a sequence of $n+2$
operators, considering only the  commutator (i) in (\compb) and (\compa), by listing the independent
strings, at the first few levels, acting on some highest-weight state
$$\eqalignT{  
& ~& \A-{\rm type} \quad {\rm (using~only }\; (i))\qquad & \B-{\rm type} \quad {\rm (using~only }\; (i))\qquad
\cr &s=1\qquad& \A_{-1}\cdots\A_{-1}\quad & \B_{n}\B_{n-1}\cdots \B_{0}\B_{-1}\cr
&s=2& \A_{-2}\A_{-1}\cdots\A_{-1}\quad & \B_{n-1}\B_{n-1}\B_{n-2}\cdots \B_{0}\B_{-1}\cr
&s=3& \A_{-3}\A_{-1}\cdots\A_{-1}\quad & \B_{n-2}\B_{n-1}\cdots \B_{0}\B_{-1}\cr
&~& \A_{-2}\A_{-2}\A_{-1}\cdots\A_{-1}\quad & \B_{n-1}\B_{n-2}\B_{n-2}\cdots \B_{0}\B_{-1}\cr}\eq$$

\subsec{A hidden $\Z_3$ exclusion principle}

%%%

Now, given that $\B\B\sim \A$, and that the $\A$ basis is ordered, we expect that the other
commutation relations $(ii)$ and $(iii)$ of
(\compb) will provide further constraints. Using the bijection just described, we would like to study the
effect of these constraints by  unraveling those  encoded
in the commutation relations $(ii)$ and $(iii)$ of (\compa). But at this point, the difference between both
versions of $(iii)$ become relevant: it would be legitimate to use the bijection only if we
had $\A^{(3)}_{n+m}=0$. However, this is never possible. The closest we can get to this is to enforce
$$ \A^{(3)}_{n+m}\sim \delta_{n+m,0}\eq$$
which holds only for the $\Z_3$ model (where $\psi_3\sim(\psi_1)^3)\sim \II$). 
But fortunately, as  far as the constraints on a basis are concerned, this delta term  is meaningless.

Now, extracting the  constraints encoded in  $(ii)$ and $(iii)$ of (\compa) when $k=3$ on a free ordered basis
simply amounts to describe the underlying $\Z_3$  quasi-particle $\A$-type  basis. And for this we can rely
on the results of [\JMb]: the $\Z_3$ quasi-particle basis is obtained by excluding, in any ordered string of
$\A$ operators, all those containing one of the following 3-strings:
$$(\A_{-(n'+1)})^{3-i}(\A_{-n'})^{i}\qquad \forall n'\geq 1,\quad (i=0,1,2)\eqlabel\parexcltr$$
Equivalently, the ordered sequences of $\A$ operators $\A_{-n'_1}\cdots \A_{-n'_p}$ are subject to 
the condition
$$n_i'\geq n_{i+2}'+2\eq$$
Now, by subtracting the staircase partition, this condition translates into 
$$n_i\geq n_{i+2}\eq$$
In other words, the next-nearest neighbors are ordered. This condition ensures that when paired two by two,
the composite
$\B$ indices are ordered:
$$n_i+n_{i+1}\geq n_{i+2}+n_{i+3}\eq$$
as expected from the identity $\B\B\sim A$.
Hence, the initially allowed step by step increases in the $\B$ mode
indices is thus finally truncated to a single step possible increase.

Summarizing these results, we have found that $\B$ strings can
only be {\it partially ordered}: strings of the form
$$\B_{-n_1}\B_{-n_2}\cdots \B_{-n_p}|\vp_q\R\eq$$ are constrained by the {\it weak-ordering
conditions}:
$$ n_i\geq n_{i+1}-1,\qquad n_i\geq n_{i+2}
\eqlabel\oror$$
In particular, the ground state among all those states containing $p$ operators $\B$ is 
$$\B_{-i}\cdots \B_{0}\B_{-1}\B_{0} \B_{-1}|\vp_q\R\eq$$
with $i=0$ if $p$ is even and $1$ otherwise.  

We  stress that these weak-ordering conditions that have just been obtained are those of a free basis. They
do not encode any $\Zt_k$ exclusion principle yet. We used a hidden $\Z_3$ $\A$-type exclusion principle only
to unravel, in a simple and elegant way, the $\B$ ordering condition. 

\subsec{The $\Zt_k$ standard basis summarized}

The analysis of the previous section  applies as well to $\B^\da$ strings, which are thus also weakly
ordered. In particular, the ground state is of the from the ground state among all those states containing $p$
is $(\B^\da_0)^p|\vp_q\R$. Note that the reason for which the ground state is not  
$\B_{i}^\da\cdots \B_{1}^\da\B_{0}^\da\B_{1}^\da\B_{0}^\da |\vp_q\R$ (with $i=0$ if $p$ is even and $1$
otherwise) is due to the highest-weight condition
$\A_{n>0}^\da|\vp_q\R=0$, in particular $\A_{1}^\da|\vp_q\R=0$.

 For mixed strings, the set of all
states with fixed charge $r+q$ in the module of highest-weight state $|\vp_q\R$ is then generated by 
$$ \B_{-n_1}\B_{-n_2}\cdots \B_{-n_p}\B^\da_{-m_1}\B^\da_{-m_2}\cdots \B^\da_{-m_{p'}}|\vp_q\R\eq$$ 
with the weak ordering condition (\oror) applying separately for the $n$'s as well as for the $m$'s, with 
$n_p\geq 1, \;
m_{p'}\geq 0$ and $p-p'=r$.  This is the standard basis of the $\Zt_k$ model.

\newsec{The spectrum of the  $\Zt_k$ model}

\subsec{The dimension of the $\Zt_k$ primary fields}

The conformal dimension of a highest-weight state is related to its charge in a way that is
fixed by the first two commutators in (\graco) and by the relation (\tem):
$$h_{\vp_q}= {q(2k-3q)\over 4k (2k+3)}\eqlabel\pridi$$
Observe that the $\Z_k$ symmetry is broken at the level of the vacua: $\vp_q^\da\not=\vp_{k-q}$ since
they do not have the same dimension.

\subsec{Reducible modules and singular vectors}

For generic values of $q$, the modules are irreducible. For $q$ half-integer, there is one
singular vector: $(\B_0^\da)^{2q+1} |\vp_q\R $.  However, when
$q$ is integer and lies in the range $0\leq q\leq k$, we  show, by construction, that there is an infinite
number of singular vectors.

The simplest way of identifying these singular vectors is to look for states whose dimensions
are related to their charge exactly as in (\pridi). In this way, we find the following two sequences of
singular vectors:
$$ \eqalign{
&  (\A_{-1})^{(2\ell+1)(k+2)+q-\ell}(\B_0^\da)^{(2\ell)(2k+3)+2q+1} \cdots
 %(\B_0^\da)^{2(2k+3)+2q+1} 
(\A_{-1})^{(k+2)+q} (\B_0^\da)^{2q+1} |\vp_q\R\cr
& (\B_0^\da)^{(2\ell+2)(2k+3)-2q-1} (\A_{-1})^{(2\ell+1)(k+2)-q-\ell-1}\cdots
 (\B_0^\da)^{2(2k+3)-2q-1} (\A_{-1})^{(k+2)-q-1}|\vp_q\R\cr}\eqlabel\sisi$$
The above constraints on $q$ follow by enforcing all these powers to be positive integers.
 These expressions are singular for all positive values of $\ell$. Each sequence can end either with a
$\B_0^\da$ or a
$\A_{-1}$ factor.
 The powers of $\A_{-1}$ can be written in a somewhat more natural way as follows:
 $$\eqalign{
 & (\A_{-1})^{(2\ell+1)(k+2)+q-\ell}= (\A_{-1})^{\frac12[2(2\ell+1)(k+3/2)+2q+1]}\cr
& (\A_{-1})^{(2\ell+1)(k+2)-q-\ell-1}= (\A_{-1})^{\frac12[2(2\ell+1)(k+3/2)-2q-1]}\cr
}\eq$$

If $q'$ stands for the charge of one of these singular vectors, the shifted value of the charge (i.e.,
$q+1/2$) is either
$ 2m(k+3/2)+q+1/2$ or
$-2m(k+3/2)-q-1/2$ (where $m$ is either $2\ell$ or $2\ell=1$), as expected from the expressions of the
$\osp(1,2)_k$ singular vectors [\ref{J.-B. Fan and M. Yu, {\it Modules over affine Lie superalgebras}, 
hep-th/9304122.}].\foot{Denote by $\J^\pm,\, J^0$ the $\su(2)$ currents -- which are also the even
currents of $\osp(1,2)$ -- and by
$j^\pm$ the fermionic generators of $\osp(1,2)$. One primary singular vector is  inherited from the
finiteness of the
$osp(1,2)$ Lie algebra representation characterized by the  $su(2)$ highest weight $q$, with highest-weight
state
$|q\R$:
$(j^-_0)^{2q+1}|q\R$. The affine singular vector is $(J^+_{-1})^{k-q+1}|q\R$. Indeed, $J^+_{-1}$ increases the
grade by one  and the weight (Dynkin label) by two. It can be decomposed in a product of fermionic
generators as
$j^+_0j^+_{-1}$;
$j^+_0$ increases the weight by one without affecting the grade while $j^+_{-1}$ increases both the grade
and the weight by one. The point is that the state reached by
$j^+_{-1}(j^+_0j^+_{-1})^{k-q}$ is not singular; the singular state is reached by the further action of
$j^+_0$.  Hence the affine singular vector can be expressed solely in terms of the $\su(2)$ generator
$J^+_{-1}$.  The correspondence with the $\Zt_k$ model is obtained as follows: the $su(2)$ weight $q$
becomes the $\Zt_k$ charge and $j^+
\sim \B$, $j^-
\sim \B^\da$, $J^+\sim \A$, $J^-\sim\A^\da$.}

The singular nature of the first two `primary' singular vectors $(\B_0^\da)^{2q+1} |\vp_q\R$ and
$(\A_{-1})^{(k+2)-q-1}|\vp_q\R$ is established in app. A.  The argument readily implies that all the
vectors (\sisi) are singular.

 The bosonic character formulae can be written down directly by subtracting and
adding successively the contribution of these singular vectors appropriately  folded in the module of
relative charge
$r$ (see e.g., [\JMa] for the corresponding expressions of the $\Z_k$ characters). These will be
presented elsewhere.

\subsec{The spectrum of fixed-charge highest-weight states and field identifications}

Having identified the parafermionic primary fields associated to the completely degenerate
representations, $\vp_q$ for $0\leq q\leq k$,  we now look for the highest-weight states 
in fixed charge modules.  This amounts to determine the possible lengths of the lowest-dimensional strings 
$\B_{-i}\cdots
\B_0\B_{-1} \B_0\B_{-1}$ $(i=0,1)$ and $(\B_0^\da)^m$ that can be applied on $|\vp_q\R$. Of course, the main
constraint comes from the two primary singular vectors. Consider first the lowest dimensional string 
containing $p$ operators
$\B$:
$$\B_{-i}\cdots \B_0\B_{-1}
\B_0\B_{-1}|\vp_q  \R \equiv |\vp_q^{(p)}\R\eq$$
Since $(\B_0\B_{-1})^{k-q+1}|\vp_q\R$ is singular, we set $p\leq 2(k-q)+1$. The dimension of these strings
is simply\foot{The condition $n_i\geq n_{i+1}-1$ has been overlooked in [\CRS] and in consequence the
lowest-dimensional $\B$ string is not written in the form $\B_{-i}\cdots\B_0\B_{-1}$. This affects the
resulting dimensions. The contact with the coset fields is done as follows: setting $p=m-q$ and writing the
fields $\vp_q^{(\pm p)}$ under the form ${\tilde \phi}_{\{q;m\}}$, their dimension can be reexpressed as
$$h_{\{q;m\}}= {q(2q-3k)\over 4k(2k+3)}-{m^2\over 4k}+\theta(m-q)\left({m-q\over
2}+{\e^{(m-q)}\over2}\right)$$
($\theta(m-q)$ being the step function).  Note that the last term in parenthesis is an integer. The bounds on
the values of
$q$ and $m$ match those induced by the coset description once the field identifications - treated below --
are considered.  There is another superficial disagreement between our results and those of [\CRS] in that the
fractional part
$1/4k$ in $B$ or
$B^\da$ modes (and
$1/k$ in the
$A$ and
$A^\da$ ones) is never written; however it is taken into account in the  computation of the dimensions.}
$$h_{\vp_q^{(p)} }= h_{\vp_q}+{p(2k-2q-p)\over 4k}+{\e^{(p)}\over 2}\eq$$
where $\e^{(p)}=1\, (0)$ if $p$ is odd (even).
On the other hand, we have
$$(\B_0^\da)^p|\vp_q\R\equiv |\vp_q^{(-p)}\R \qquad 0\leq p\leq 2q\eq$$
and
$$h_{\vp_q^{(-p)}}= h_{\vp_q}+{p(2q-p)\over 4k}\eq$$

This analysis must be supplemented by the determination of the field identifications. In the present  case,
this amounts to identify fields having the same conformal dimension but charge differing by $2k$. We thus look
for fields related by
$p+m=2k$: setting
$p=2k-x$ and using the above bounds, we find $2q-1\leq x\leq 2q$. The charge consideration  leads thus to the
following two candidate field identifications
$$\eqalignT{
&{(i)}:\quad & p=2k-2q+1\,, \qquad & m=2q-1\cr
&{ (ii)}:\quad & p=2k-2q\,, \qquad & m=2q\cr
% &{\rm (iii)}:\quad & p=2k-2q-1\,, \qquad & m=2q+1\cr
}\eq$$
In both cases, the equality of the conformal dimensions is  easily established, leading to the
identifications:
$$\B_{-1} (\B_0\B_{-1})^{k-q}|\vp_q\R\sim (\B_0^\da)^{2q-1}|\vp_q\R\eq$$
and
$$(\A_{-1})^{k-q}|\vp_q\R\sim (\A_{0}^\da)^{q}|\vp_q\R\eq$$
(which the known $\Z_k$ one).
For $q=0$ these identifications imply that 
$$\vp_0^{(2k+1)}\sim \vp_0^{(1)}\qquad \vp_0^{(2k)}\sim \vp_0^{(0)}\eq$$
(the first relation being in fact a consequence of the second one).
For the set $\{\vp_0^{(p)}\}$, $p$ should thus be restricted to $0\leq p\leq 2k-1$. Taking into account the
field identifications, we thus end up with the following candidate, denoted   $\Gamma_{(k)}$, for the set of
fields associated to fixed-charge highest-weight states:
$$\Gamma_{(k)}= \left\{\matrix{\vp_q^{(p)}\;, \qquad &0\leq p \leq {\rm min }\; (2k-1,2k-2q+1)\cr
\vp_q^{(-p)}\;, \qquad &1\leq p \leq {\rm max }\;(1,2q-2)\cr}\right.\eq$$
For instance, for $k=2$, it reads
$$\Gamma_{(2)}= \left\{ \vp_0^{(0,1,2,3)}, \, \vp_1^{(0,1,2,3)}, \, \vp_2^{(0,1)}, \,
\vp_2^{(-1,-2)}\right\}\eq$$
Although most of these fields in $\Gamma_{(k)}$ are $\B$ descendants of $\vp_q$, some are not, e.g.,
$\vp_2^{(-1,-2)}$ in $\Gamma_{(2)}$.  However, the mere existence of a $\B$-type quasi-particle basis
requires, from the onset, that all highest-weight states in a fixed charged sector  be expressible as 
$\B$-descendants of $|\vp_q\R$. In fact this is indeed so: the states $|\vp_q^{(-p)}\R$ can be written
as\foot{Hence $|\vp_q^{(-p)}\R$ is a descendant of $|\vp_q^{(2k-p)}\R$ but not a $\B$ descendant; it can be
reached by acting with an appropriate combination of an  $\O$-type operator.}
$$|\vp_q^{(-p)}\R=(\B_0^\da)^{p}|\vp_q\R\sim (\B_{-1})^{2q-p} (\B_0\B_{-1})^{k-q}|\vp_q\R\eq$$
The dimensions are easily checked to match and the charges ($q-p$ on the left hand side and $2k-p+q$ on the
right hand side) differ by $2k$.
We thus end up with the following reduced set $\tilde{\Gamma}_{(k)}$
$$\tilde{\Gamma}_{(k)}= \left\{ \vp_q^{(p)}\;|\; 0\leq p \leq {\rm min }\; (2k-1,2k-2q+1)\right\}\eq$$
whose cardinality is $k^2+3k$.
Of course, the set $\tilde{\Gamma}_{(k)}$ could equally well be described solely in terms of $\B_0^\da$ and
$\B_{-1}^\da$ descendants.

Let us now conclude this section with a remark concerning the limit on the number of parafermions which is
induced by the primary singular vectors (and the field identifications). Recall that the parafermionic fields
are in correspondence with the states obtained by the lowest-energy strings of parafermionic modes on the
vacuum, i.e.,
$$\psi_r(0)|0\R \sim \B_{-i}\cdots \B_0\B_{-1}|0\R \qquad i=0\;  (1)\quad {\rm if} \; 2r\; {\rm
is~even~(odd)}\eq$$ The singular vector $ (\A_{-1})^{k+1}|0\R$ yields an upper bound on $2r$:
$2r<k+1$.\foot{As already pointed out, there is  a more restrictive bound -- preventing  $(\B_0\B_{-1})^k
|0\R$ and 
$\B_{-1}(\B_0\B_{-1})^k |0\R$ to correspond to novel fields -- which is induced by the field identification
$(\A_{-1})^{k}|0\R\sim |0\R$. But this is not our main point here.} However, we can similarly generate the
parafermions from the
$\B^\da$ states as
$$\psi^\da_r(0)|0\R\sim \B^\da_{-i}\cdots \B_0^\da\B_{-1}^\da|0\R \qquad i=0\;  (1)\quad {\rm if} \;2r\; {\rm
is~even~(odd)}\eq$$
(and not $(\B^\da_{0})^r|0\R $ since this is a descendant of a singular state). At first sight there
seems to be no constraint on the length of such strings. It is however also a consequence of the
singular vector $ (\A_{-1})^{k+1}|0\R$ appropriately folded.  Take for simplicity the case where $r$ is
integer, so that $\psi^\da_r(0)|0\R\sim (\A^\da_{-1})^{r}|0\R$. The singular vector $ (\A_{-1})^{k+1}|0\R$ can
be folded in the $-2r$ charged sector by the action of the  operator $(\A^\da_0)^{r+k+1}$. Note however that
the resulting state has level $k+1$. Hence, it can only subtract the state $(\A^\da_{-1})^{r}|0\R$ for
$r=k+1$. We thus recover that same constraint as previously.

\subsec{The $\Zt_1$ model}

% \n $\bullet$ Compare the $\Zt_1$ spectrum 
% with the ${\cal M}(3,5)$ model! 

For $k=1$, the central charge of the $\Zt_k$ model reduces to $-3/5$, which is that of the
${\cal M}(3,5)$ Virasoro minimal model.  
The spectrum of the ${\cal M}(3,5)$ model reads $$h_{11}=0\,,\qquad  h_{12}=-1/20\,,\qquad 
h_{13}= 1/5\,,\qquad h_{14}=3/4\eq$$
This match precisely that of our graded parafermionic theory, for which the distinct fixed charged
highest-weight states are
$|0\R\,, \B_{-1}|0\R\,,|\vp_1\R \,,  \B_{-1} |\vp_1\R$
of respective dimension $0,\,3/4,\, -1/20,\,1/5$. 
We thus have the following correspondence:
$$\eqalignD{
& |\phi_{11}\R\sim
|0\R \,,\qquad  &|\phi_{14}\R\sim
\B_{-1}|0\R=|\psi_{\frac12}\R\cr
&|\phi_{12}\R\sim
|\vp_1\R  \,,\qquad  &|\phi_{13}\R\sim
\B_{-1}|\vp_1\R \cr}\eq$$

Note that this is the only $\Zt_k$ model
that also has a description in terms of a Virasoro minimal model. Moreover, none of the $\Zt_k$ models has
a description as a superconformal minimal model.

\subsec{The Neveu-Schwarz  and 
Ramond sectors}

Given that the $\B$ operators are fermionic, we might wonder whether there is a Neveu-Schwarz (NS) and a
Ramond (R) sector in the $\Zt_k$ models. Actually, the $\B$'s are also parafermionic and in parafermionic
theories, the sectors are fixed by the charge of the state on which the operators act, i.e., that fixes the
fractional part of the modes. In the
$\Z_2$ model (equivalent to the Ising model), this fractional part either vanishes or is half-integer. This
suggests that the two extremal values of $q$, namely $0$ and $k$, could be though as the realization of the R
and NS sectors respectively since the fractional parts differ by $1/2$. In the $\Zt_1$ model -- which is quite
similar to a free fermion theory as it will be shown below --, these are the
only two sectors. However this analogy is not very compelling 
since the fractional part of successive modes, in
each sector,
 differ by $1/2$.

\newsec{The $\Z_k$  quasi-particle basis}

% \subsec{The $\Zt_k$ exclusion principle}

The $\Zt_k$ quasi-particle basis is also spanned by strings of a single type of $\Zt_k$ parafermions, say the
$\B$ modes:
$$\B_{-n_1}\B_{-n_2}\cdots \B_{-n_p}|\vp_q\R\qquad {\rm with}\quad 
 n_i\geq n_{i+1}-1\, ,\qquad n_i\geq n_{i+2}\, ,\qquad n_p\geq 1 \eqlabel\ororf$$
In addition, we need to take into account the singular vector $(\A_{-1})^{k-q+1}|0\R=
(\B_0\B_{-1})^{k-q+1}|0\R$, which implies that either 
$$n_{p-2k+2q}\geq 2\qquad {\rm or } \qquad   n_{p-2k+2q+1}\geq 1\eqlabel\bdry$$ Indeed,
there are two ways to terminate a sequence of pairs $\B_0\B_{-1}\cdots\B_0\B_{-1}$: one can either act with
$\B_{-2}$  or with $\B_{-1}\B_{-1}$ on the left. The constraint (\bdry) is thus our boundary condition.

Consider now the constraint that comes from the $\Z_k$ invariance of the theory. Actually, in terms of the
$\B$ modes, it is rather a $\Z_{2k}$ invariance, namely $(\psi_{\frac12})^{2k}\sim \II$. This implies that
all weakly ordered $\B$-strings containing $2k$ elements are not independent.  In fact, there is one
relation at each level. This linear relation forces the exclusion of one $2k$-string at each level.  The
chosen excluded states need to have indices `as equal as possible' in order for the exclusion of the selected
$2k$-strings within longer strings to do not generate contradictions. This governs the following choice of
{\it excluded}
$2k$-strings:
$$\eqalignD{
& (\B_{-(n-1)}\B_{-n})^{k-j} (\B_{-n}\B_{-n})^{j}  \qquad  &j=1,\cdots, k\cr
& (\B_{-n}\B_{-(n+1)})^{k-j} (\B_{-n}\B_{-n})^{j}  \qquad  & j=1,\cdots, k\cr}\eq$$
Phrased differently, the excluded states are those containing $2k$-strings of the form
$\B_{-n_i}\B_{-n_2}\cdots
\B_{-n_{i+2k-1}}$ satisfying
$$\eqalignT{
& {\rm {\bf  Excluded}:}\quad  & n_i = n_{i+2k-1}-1  & \cr &  & {\rm or}  & \cr
& ~ &n_i = n_{i+2k-1} \quad &{\rm and}
\quad |n_j-n_\ell|\leq 1 \cr &~ &  &
\forall j,\ell\in [i,\cdots,i+2k-1]\cr}\eq$$
Instead of characterizing those states that are excluded, one can formulate the condition in terms of allowed
states, which are those general $\B$ strings of the form $\B_{-n_1}\B_{-n_2}\cdots \B_{-n_{p}}$
satisfying
$$\eqalignD{
& {\rm {\bf Allowed}:}\quad &n_i \geq  n_{i+2k-1} +1\cr & & {\rm or} \cr 
&~& n_i = n_{i+2k-1} \quad {\rm and}
\quad n_{i+1} \geq  n_{i+2k-2}+2 \cr}\eqlabel\allo$$ 
for all values of $i\leq p-2k+1$.

To make these conditions clearer, we provide some examples. For simplicity, we will again write a sequence
$\B_{-n_1}\B_{-n_2}\cdots \B_{-n_{2k}}$ under the form $(n_1,\cdots n_{2k})$. Consider first the case
$k=2$. The first four lowest dimensional excluded $4$-strings are 
$$(0101)\qquad (1101)\qquad (1111) \qquad (1211)\eq$$
For the first one, $n_1< n_4$. In the other three cases, we have $n_1=n_4$ but all the differences are
$0,\pm1$; hence these are also excluded. By adding to these four excluded $4$-strings the
partition $(n,n,n,n)$ for any integer $n$, we generated all excluded 4-strings. Not that at `level' 4
there is one allowed string with $n_1=n_4$, which is $(1201)$; it indeed satisfies $n_2\geq n_3+2$.

As a second example, consider $k=3$. The lowest dimensional excluded strings are
$$(010101)\qquad (110101)\qquad (111101) \qquad (111111) \qquad (121111)\qquad (121211)\eq$$
and again there are 6-strings with `level' 4 and 5 that are allowed, namely $(120101)$ and $(121101)$ which
are both characterized by the `internal difference two condition': $n_2\geq n_5+2$.

Let us now take the case $k=1$: the lowest excluded 2-strings are $(01)$ and $(11)$. In that case, the second
condition in (\allo) is not applicable and thus the exclusion principle boils down to $n_i\geq n_{i+1}+1$. 
This is actually the sole condition since it is  stronger than the weak-ordering conditions (\oror).  We thus
end up with a purely fermionic Fock basis for the $\Zt_1$ model.

Another example is reported in app. B.

\newsec{Conclusions}

The main result of this work is the presentation of the two bases of states for the $\osp(1,2)_k/\uh(1)$
parafermionic model.  At first we have obtained the weak
ordering condition for a free $\B$ basis by means of a bijection with $\Z_3$ $\A$-strings. The same ordering
applies to $\B^\da$-strings. Since there is no mixed ordering, this fixes the standard basis.  On the other
hand, the quasi-particle basis is obtained from an ordered $\B$ basis
on which we  implement the
$\Z_{2k}$ invariance; the effect of this constraint is captured by an effective
$\Zt_k$ exclusion principle.  

As a side result, we have clarified: (1) some aspects of the representation
theory of the
$\Zt_k$ models by displaying the full set of singular vectors in closed form and (2) the spectrum of
the theory, including the field identifications. 

Of course, the immediate extension of this work is to write down the bosonic and fermionic-type character
formulae. The later is somewhat complicated by the fact that we do not count genuine restricted partitions.
But apart from this, we can identify other natural avenues for future work.

The $\osp(1,2)_k/\uh(1)$ model is the simplest $\gh/ \uh(1)^r$ model for which $g$ is a super Lie algebra. It
would be interesting to consider other examples of super models of that sort. However, the
$\osp(1,2)_k/\uh(1)$ model has a more naive generalization which could be formulated as follows. For any
(nonsuper) Lie algebra $g$, we could extend the $\gh/ \uh(1)^r$ theory by adding to the set of fundamental
parafermions $\psi_{1}^{\alpha_i}$,  where
$\alpha_i$ is a simple root, their `square roots'
$\psi_{\frac12}^{\alpha_i}$. That would define a candidate graded version of the  $\gh/ \uh(1)^r$ model.
Whether this actually defines a model satisfying associativity and whether it is related to a specific
parafermionic theory based on a super affine  Lie algebra  remains to be checked. 
 
% For the $\osp(1,2)_k/\uh(1)$ model, the quasi-particle 
% basis has a hidden  $\Z_3$ exclusion
% principle implemented over a $\Z_{2k}$ one. 
% It is natural to look for the potential
% existence of generalized parafermionic models 
% with a hidden $Z_{p>3}$ exclusion principle over which a
% generic $\Z_pk$ invariance is enforced. 

Another natural future problem is to study the integrable 
deformations of the $\osp(1,2)_k/\uh(1)$ model and their flows.  In relations with the results of the
present paper, it will be of interest to identify the precise integrable perturbation associated to the
quasi-particle basis obtained here.

We end with an  observation that is potentially interesting from the point of view of
condensed matter applications: the
$\Zt_k$ models in the limit $k\rw \y$ provide a well-defined limiting approach to a $c=0$ theory.

% the occurence of the...one is somewhat strange: 
% the natural guess would be that there is a
% $\Z_2$ invariance (associated to the $\Z_2$ grading)
%  over which we enforce the standard
% $\Z_k$ one.

%  \appendix{A}{An effective $\Z_3$ exclusion principle and the weak ordering condition} 

%----- comment on:
% The $\Zt_k$ ordering  turns out to be weaker and governed by
% a hidden $\Z_3$ exclusion principle. 
% This hidden structure can be unraveled by comparing the $\B\,\B$ and
%  $\B\,\B^{(2)}$  commutation relations 
% in the $\Zt_k$ model with the $\A\,\A$ and
% $\A\,\A^{(2)}$  commutation reations in the
% $\Z_3$ model. 

%=====================================

\appendix{A}{The `primary' $\Zt_k$ singular vectors}

In this appendix, we demonstrate that the two vectors $(\A_{-1})^{k-q+1}|\vp_q\R$ and
$(\B^\da_{0})^{2q+1}|\vp_q\R$ are singular. In the first case, we first establish the following relations
$$\eqalignD{
&\A_p (\A_{-1})^{n}|\vp_q\R =0  &\qquad p\geq 0\,,\qquad \forall n\cr
&\B_p (\A_{-1})^{n}|\vp_q\R=0   &\qquad p\geq 0\,,\qquad \forall n\cr
&\B^\da_p (\A_{-1})^{n}|\vp_q\R =0 &\qquad p\geq 1\,,\qquad \forall n\cr}\eq$$
in this order, the first (second) relation being used in the proof of the second (third) one. All
these relations are easily verified for $n=1$ by a direct application of the generalized commutation
relations. The general case is proved by induction.  At this point, the value of $n$ is still free. To verify
the remaining condition,
$\A^\da_p (\A_{-1})^{n}|\vp_q\R $, we introduce a variant of the second  commutator in (\graco):\foot{In this
way we do not have to know the commutator of $\O_m^{(1)}$ with the $\A$ modes as well as its zero-mode
eigenvalue on a generic state.}
$$ \sum_{l=0}^{\infty} C^{(l)}_{-2/k} \left[ \A_{n-l}
\A^\da_{m+l} - \A^\da_{m-l} \A_{n+l}
 \right] \,\phi_{q}(0)  
 =  
\left({q\over k}+n\right)\delta_{n+m,0} \,\phi_{q}(0) 
\eq$$
This allows us to show readily that
$$ \A^\da_p (\A_{-1})^{n}|\vp_q\R =0  \qquad p\geq 2\,,\qquad \forall n\eq$$
again by induction. The final condition is
$$ \A^\da_1 (\A_{-1})^{n}|\vp_q\R =  {(n-1)\over k}(q+n-1-k)(\A_{-1})^{n-1}|\vp_q\R\eq$$
whose vanishing is verified only for $n=k-q+1$.

For the vector $(\B^\da_{0})^{2q+1}|\vp_q\R$, we proceed similarly, obtaining first
$$\eqalignD{
&\A^\da_p (\B^\da_{0})^{n}|\vp_q\R =0 \qquad  &p\geq 1\,,\qquad \forall n\cr
&\B^\da_p (\B^\da_{0})^{n}|\vp_q\R=0  \qquad  &p\geq 1\,,\qquad \forall n\cr
}\eq$$
It remains to consider the $\B_p$ condition (which implies the $\A_p$ one) and for this we use
$$ \sum_{l=0}^{\infty} C^{(l)}_{-2/k} \left[ \B_{n-l}
\B^\da_{m+l} +\B^\da_{m-l} \B_{n+l}
 \right] \,\phi_{q}(0)  
 =  
\left({q\over 2k}+n\right)\delta_{n+m,0} \,\phi_{q}(0) 
\eq$$
Again the case $p\geq 1$ is direct
$$\B_p (\B^\da_{0})^{n}|\vp_q\R=0  \qquad p\geq 1\,,\qquad \forall n\eq$$
while imposing the $\B_0$ highest-weight condition leads to
$$\B_0 (\B^\da_{0})^{n}|\vp_q\R= {n\over 4k}(2q-n-1) (\B^\da_{0})^{n}|\vp_q\R=0\eq$$ which is zero if and
only if
$n=2q+1$.

This argument extends directly to the general case: it relies simply on the fact the considered power of
$\A_{-1}$ or $\B^\da_0$ act on a highest-weight state and that the highest-weight state condition forces a
precise relation between its  dimension and its charge. 

\appendix{B}{Standard vs quasi-fermionic basis: counting the first few states in chargeless vacuum module of
$\Zt_3$ model}

To illustrate the construction of a module level-by-level by using the two bases constructed in this work, we
write down the different states that occur at level $s=5$ in the vacuum module of zero relative charge in
the $k=3$ model.

\subsec{The standard $\Zt_3$ basis ($q=0,\, r=0,\,k=3,\, s=5$)}

Zero charge states
in the vacuum module are generated by 
$$ \B_{-n_1}\B_{-n_2}\cdots \B_{-n_p}\B^\da_{-m_1}\B^\da_{-m_2}\cdots \B^\da_{-m_{p}}|0\R\eqlabel\seqq$$ 
with the weak ordering conditions
$$ n_i\geq n_{i+1}-1\, ,\quad n_i\geq n_{i+2}\,,\quad n_p\geq 1\,,\quad m_i\geq m_{i+1}-1\,,\quad m_i\geq
m_{i+2}\, ,\quad m_p\geq 0
\eqlabel\ororo$$ for any integer $p$. The level of such  a state is determined by the conformal dimension of
the string, which itself depends upon the fractional value of its modes. However, it is simple to check that
when the number of $\B$ and $\B^\da$ factors is the same, the fractional parts add up to zero. Therefore the
level
$s$ is given by
$$s=\sum_i( n_i+m_i)\eq$$ 

For the vacuum module of the $k=3$ model at level $s\leq 5$, the following singular vectors 
need to be taken into account:
$$\B_0^\da|0\R\,,\qquad (\A_{-1})^4|0\R\,,\qquad (\A_{-1})^5\B_0^\da|0\R\eq$$
We must subtract the contribution of the first two and add up that of the third one.  The effect of the first
one can be implemented by enforcing $m_p\geq 1$ in (\seqq). With this constraint, we find 18 states at
level 5:
$$\eqalignD{
& \B_{-4}\B^\da_{-1}\qquad & \B_{-3}\B^\da_{-2}\cr
& \B_{-2}\B^\da_{-3}\qquad & \B_{-1}\B^\da_{-4}\cr
& \B_{-3}\B_{-1}\B^\da_{0}\B^\da_{-1}\qquad &\B_{-2}\B_{-2}\B^\da_{0}\B^\da_{-1}\cr
& \B_{-2}\B_{-1}\B^\da_{-1}\B^\da_{-1}\qquad &\B_{-1}\B_{-2}\B^\da_{-1}\B^\da_{-1}\cr
& \B_{-1}\B_{-1}\B^\da_{-2}\B^\da_{-1}\qquad &\B_{-1}\B_{-1}\B^\da_{-1}\B^\da_{-2}\cr
& \B_{0}\B_{-1}\B^\da_{-1}\B^\da_{-3}\qquad &\B_{0}\B_{-1}\B^\da_{-2}\B^\da_{-2}\cr
& \B_{-2}\B_{0}\B_{-1}\B^\da_{-1}\B^\da_{0}\B^\da_{-1}\qquad &
\B_{-1}\B_{-1}\B_{-1}\B^\da_{-1}\B^\da_{0}\B^\da_{-1}\cr
& \B_{-1}\B_{0}\B_{-1}\B^\da_{-2}\B^\da_{0}\B^\da_{-1}\qquad &
\B_{-1}\B_{0}\B_{-1}\B^\da_{-1}\B^\da_{-1}\B^\da_{-1}\cr
& \B_{-1}\B_{-1}\B_{0}\B_{-1}\B^\da_{0}\B^\da_{-1}\B^\da_{0}\B^\da_{-1}\qquad &
\B_{0}\B_{-1}\B_{0}\B_{-1}\B^\da_{-1}\B^\da_{-1}\B^\da_{0}\B^\da_{-1}\cr}\eq$$

We now subtract those states that are descendants of the singular vector $(\A_{-1})^4|0\R$. Its lowest
dimensional descendant of charge zero is  $(\B_0^\da)^8 (\A_{-1})^4|0\R$, which occurs at level 4. At level 5,
there are four descendants, namely:
$$\eqalignD{
& \B_{-1}^\da(\B_0^\da)^7(\A_{-1})^4|0\R\,,\qquad &\B_0^\da\B_{-1}^\da(\B_0^\da)^6(\A_{-1})^4|0\R\cr
& (\B_0^\da)^9 \B_{-1}(\A_{-1})^4|0\R\,,\qquad &(\B_0^\da)^{10} \B_0\B_{-1}(\A_{-1})^4|0\R\cr}\eq$$
Finally, because $(\A_{-1})^5\B_0^\da|0\R$ is a descendant of both $\B_0^\da|0\R$ and $ (\A_{-1})^4|0\R$, we
need to add the contribution of the zero-charge descendants of $(\A_{-1})^5\B_0^\da|0\R$. At level 5, there is
a single such state:
$(\B_0^\da)^9 (\A_{-1})^5|0\R$. The total number of states at level 5 is thus $18-4+1=15$.

\subsec{The quasi-particle $\Zt_3$ basis ($q=0,\, r=0,\,k=3,\, s=5$)}

In that case, we need to consider $\B$-strings of length $p=m2k$ subject, in addition to the weak
ordering conditions (\oror), to the $\Zt_3$ exclusion principle. 

At first, let us compute the level of a $\B$-string of length $p$ acting on a state of charge $q$. The
fractional part $F$ is found to be
$$F= {p(2q+p)\over 4k}= m(q+mk) \quad {\rm when}\quad p=m2k\eq$$
For $q=0$ and $k=3$, the contribution of the fractional part to the conformal dimension is thus $-3m^2$, i.e.,
$$s=- 3m^2 +\sum_{i=1}^{6m} n_i \eq$$

The contributing 6-strings at level 5 are:
$$\eqalignD{
&\B_{-5}\B_{-1}\B_{0}\B_{-1}\B_{0}\B_{-1}\qquad & \B_{-4}\B_{-2}\B_{0}\B_{-1}\B_{0}\B_{-1}\cr
&\B_{-4}\B_{-1}\B_{-1}\B_{-1}\B_{0}\B_{-1}\qquad & \B_{-3}\B_{-3}\B_{0}\B_{-1}\B_{0}\B_{-1}\cr
&\B_{-3}\B_{-2}\B_{-1}\B_{-1}\B_{0}\B_{-1}\qquad & \B_{-3}\B_{-1}\B_{-2}\B_{-1}\B_{0}\B_{-1}\cr
&\B_{-3}\B_{-1}\B_{-1}\B_{-1}\B_{0}\B_{-1}\qquad & \B_{-2}\B_{-3}\B_{-1}\B_{-1}\B_{0}\B_{-1}\cr
&\B_{-2}\B_{-2}\B_{-2}\B_{-1}\B_{0}\B_{-1}\qquad & \B_{-2}\B_{-2}\B_{-1}\B_{-2}\B_{0}\B_{-1}\cr
&\B_{-2}\B_{-2}\B_{-1}\B_{-1}\B_{-1}\B_{-1}\qquad & \B_{-2}\B_{-1}\B_{-2}\B_{-1}\B_{-1}\B_{-1}\cr}\eq$$
There are also three contributing 12-strings:
$$\eqalign{
&\B_{-3}\B_{-3}\B_{-2}\B_{-1}\B_{-2}\B_{-1}\B_{-2}\B_{-1}\B_{0}\B_{-1}\B_{0}\B_{-1}\cr
&\B_{-3}\B_{-2}\B_{-3}\B_{-1}\B_{-2}\B_{-1}\B_{-2}\B_{-1}\B_{0}\B_{-1}\B_{0}\B_{-1}\cr
&\B_{-2}\B_{-3}\B_{-2}\B_{-2}\B_{-2}\B_{-1}\B_{-2}\B_{-1}\B_{0}\B_{-1}\B_{0}\B_{-1}\cr}\eq$$
At level 5, there are no contributing longer strings. Therefore the total number of states is $12+3=15$.

\vskip0.3cm
\centerline{\bf Acknowledgment}

We thank IPAM and the organizers of the CFT 2001 semester for their hospitality while this work was
being done.  We thank H. Saleur and R. Kedem for useful discussions. NSERC is acknowledged for the
financial support.

\vskip0.3cm

%\vfill\eject
\centerline{\bf REFERENCES}
%\vskip 0.5cm
\immediate\closeout\refs \vskip 0.5cm
  \message{References}\input references
\vfill\eject

\end

retrait de la section:

\subsec{Reviewing the $\Z_k$ basis}

Before presenting the $\Zt_k$ quasi-particle basis, let us review the usual $\Z_k$ one [\JMb]. It is
expressed solely in terms of  the  $\A$ modes, so that the constraint
of a fixed value for the difference on the number of
$\A$ and
$\A^\da$ operators in a module of fixed charge (namely, the first condition in (\mixst)) gets replaced by
the conservation of the number of
$\A$ operators modulo
$k$.
A spanning set of states in  the highest-weight module of 
$|\varphi_\ell\R$ of relative charge
$2r$ is provided by the set of ordered strings
$$\A_{-n_1}\A_{-n_2}\cdots \A_{-n_m}|\varphi_\ell\R\qquad \quad {\rm with}\quad 
n_i\geq n_{i+1}\geq 1\eqlabel\ferbas$$  
with $m=r$ mod $k$. This must be supplemented by the `boundary condition'
$$n_{m-k+\ell}\geq 2\eqlabel\ferbass$$
which keeps track of the singular vector $(\A_{-1})^{k-\ell+1}|\varphi_\ell\R$, by ensuring that
the
$k-\ell+1$-th operator counted from the right is not
$\A_{-1}$. In addition, we impose the following selection rules: all ordered sequences
of
$\A$ operators that contain any one of the following $k$-string
$$(\A_{-(n+1)})^{k-i}(\A_{-n})^{i}\qquad (i=0,\cdots , k-1)\eqlabel\parexcl$$
must be forbidden.  This can be viewed as a generalized exclusion principle [\JMb]. 
As
stressed in App. C of [\JMb], it is an important property of these excluded
$k$-strings that they have indices `as equal as possible'. 

 Excluding the
states (\parexcl) amounts to enforce (\restru).\foot{The interpretation of this condition in terms of
an exclusion principle seems to go back to [\ref{J. Lepowsky and R.L. Wilson, Proc. Nat. Acad. Sci. USA {\bf
78} (1981) 7254.}\refname\LW].  A more recent discussion can be found in [\ref{C. Dong and J.
Lepowsky {\it Generalized vertex algebras and relative vertex operators}, Birkha\"user, 1993.}\refname\DL]. 
For another formulation of
the generalized  exclusion principle as a constraint on a basis of states, see [\ref{K.  Schoutens, Phys.
 Rev. Lett. {\bf 79} (1997) 2608; P. Bouwknegt. and K. 
 Schoutens, Nucl. Phys. 
 {\bf B 547} (1999) 501.}\refname\bouschou]. Also, the connection between such type of  exclusion
principle and the Haldane's generalization [\ref{F.D.M. Haldane,  Phys. Rev. Lett. {\bf 67} (1991) 
937.}\refname\hal] is clarified in [\ref{A. Berkovich and B.M. McCoy, 
%{\it The universal chiral partition function for
% exclusion statistics}, hep-th/9808013.}
in {\it Statistical physics at the eve of the 21st century}, 
Series on Adv. in Stat. Mech., vol 14, ed. by M.T Batchelor and L.T. White, World Scientific, 1999.}].}  This
basis turns out to be equivalent to the Lepowsky-Primc basis (with is formulated differently however -- cf.
[\LP], section 6) and it leads to the Lepowsky-Primc character formula [\LP].

%%%

%%%%%%%%%%%%%%%%%%%
Compare the last commutator in
(\graco), which we rewrite for convenience, with its $\A$ analogue:
$$\eqalign{
 & 
 \sum_{l=0}^{\infty} C^{(l)}_{1+1/2k} \left[ \B_{n-l}
\B_{m+l} - \B_{m+1-l} \B_{n-1+l}
 \right] \phi_{q}(0)  =   0 \cr
   & \sum_{l=0}^{\infty} C_{2/k}^{(l)} 
  \left[ \A_{n-l}\A_{m+l}-\A_{m-l}\A_{n+l} \right]
 \phi_{q}(0) =0 \cr}\eqlabel\compa$$
It is readily observed that the mode indices on the second term of
the
$\B$ commutator are shifted by
$+1$ and
$-1$ when compared to the corresponding
$\A$ terms.  If the $\A$ commutator can be used to 
 eliminate  a term of the form $\A_{0}\A_{-1}$ acting on a highest weight, and ultimately, to order
all the modes in a $\A$ string, the corresponding $\B$ commutator does not allow us to eliminate the
state $\B_0\B_{-1}$.  (Clearly, the fourth commutator in (\graco) is of no help in that regard since it  only
shows that
$\B_0\B_{-1}\propto\A_{-1}$.)  However, the  increase in the value of the indices from left to right is
strongly constrained: for instance, all the states 
$\B_n\B_0\B_{-1}|\vp_q\R$ can be shown to vanish unless
$n\leq -1$.